\documentclass[12pt]{article}
\usepackage{epsfig}
\usepackage{psfrag}
\usepackage{latexsym}
\usepackage{indentfirst}
\usepackage{fancyhdr}
\usepackage{amsmath,amssymb,amsfonts}
\usepackage{cite}
\usepackage{bbold}
\usepackage{color}
\usepackage{graphicx}
\usepackage[center,footnotesize,hang]{subfigure}
\usepackage{url}

\makeatletter

\newcommand{\Rmnum}[1]{\expandafter\@slowromancap\romannumeral #1@}
\makeatother

\def\be{\begin{equation}}
\def\ee{\end{equation}}
\def\bc{\begin{center}}
\def\ec{\end{center}}
\def\bea{\begin{eqnarray}}
\def\eea{\end{eqnarray}}

\catcode`@=11
\def\marginnote#1{}
\newcount\hour
\newcount\minute
\newtoks\amorpm
\hour=\time\divide\hour by60
\minute=\time{\multiply\hour by60 \global\advance\minute by-\hour}
\edef\standardtime{{\ifnum\hour<12 \global\amorpm={am}%
        \else\global\amorpm={pm}\advance\hour by-12 \fi
        \ifnum\hour=0 \hour=12 \fi
        \number\hour:\ifnum\minute<10 0\fi\number\minute\the\amorpm}}
\edef\militarytime{\number\hour:\ifnum\minute<10 0\fi\number\minute}
\def\draftlabel#1{{\@bsphack\if@filesw {\let\thepage\relax
   \xdef\@gtempa{\write\@auxout{\string
      \newlabel{#1}{{\@currentlabel}{\thepage}}}}}\@gtempa
   \if@nobreak \ifvmode\nobreak\fi\fi\fi\@esphack}
        \gdef\@eqnlabel{#1}}
\def\@eqnlabel{}
\def\@vacuum{}
\def\draftmarginnote#1{\marginpar{\raggedright\scriptsize\tt#1}}
\def\draft{\oddsidemargin 0.0truein
        \def\@oddfoot{\sl preliminary draft \hfil
        \rm\thepage\hfil\sl\today\quad\militarytime}
        \let\@evenfoot\@oddfoot \overfullrule 3pt
        \let\label=\draftlabel
        \let\marginnote=\draftmarginnote
   \def\@eqnnum{(\theequation)\rlap{\kern\marginparsep\tt\@eqnlabel}%
\global\let\@eqnlabel\@vacuum}  }
\catcode`@=12

\begin{document}
\title{\bf On determination of the geometric cosmological constant from the OPERA experiment of superluminal neutrinos}

\author{Mu-Lin Yan\footnote{E-mail address: mlyan@ustc.edu.cn}, Sen Hu\footnote{E-mail address: shu@ustc.edu.cn}, Wei Huang\footnote{E-mail address: weihuang@mail.ustc.edu.cn}, Neng-Chao Xiao\footnote{E-mail address: ncxiao@ustc.edu}
\\Wu Wen-Tsun Key Lab of Mathematics of Chinese Academy of Sciences\\
Department of Modern Physics and School of Mathematical Sciences\\
University of Science and Technology of China, Hefei, Anhui 230026, China
}
\maketitle

\begin{abstract}
The recent OPERA experiment of superluminal neutrinos has deep consequences in cosmology. In cosmology a fundamental constant is the cosmological constant. From observations one can estimate the effective cosmological constant $\Lambda_{eff}$ which is the sum of the quantum zero point energy $\Lambda_{\mbox{dark energy}}$ and the geometric cosmological constant $\Lambda$. The OPERA experiment can be applied to determine the geometric cosmological constant $\Lambda$. It is the first time to distinguish the contributions of $\Lambda$ and $\Lambda_{\mbox{dark energy}}$ from each other by experiment. The determination is based on an explanation of the OPERA experiment in the framework of Special Relativity with de Sitter space-time symmetry.


\end{abstract}

\newpage

\section{Introduction}

\noindent The recent OPERA experiment shows evidence of superluminal behavior for muon neutrinos $\nu_{\mu}$ \cite{OPERA}. The arrival time of the neutrinos $\nu_{\mu}$ with average energy of 17 GeV is shorter than that of photons by
$\delta t = (60.7 \pm 6.9_{stat} \pm 7.4_{sys})$ ns. In terms of speed the neutrinos went faster by
an relative amount

$$\delta v_{\nu} = \frac{v_{\nu} - c}{c} = (2.48 \pm 0.28_{stat} \pm 0.30_{sys}) \times 10^{-5}$$

\noindent with significance level of 6$\sigma$.

This puzzle is understood in the framework of Special Relativity with de Sitter space-time symmetry\cite{Yan1}. The de Sitter spacetime appeared as a solution of the Einstein's equations. When the radius goes to infinity it approximates the Minkowski spacetime. The OPERA experiment showed that we should use Special Relativity with de Sitter spacetime geometry rather than the usual Minkowski spacetime symmetry. The radius of the pseudo-sphere of de Sitter spacetime can be estimated from the experiment \cite{Yan1}.

 In this paper we shall derive another consequence of the OPERA experiment to cosmology, i.e. to determine the geometric cosmological constant.

 Based on the standard model of cosmology $\Lambda$CDM \cite{CDM} the universe is composed of atoms, dark matter and dark energy. The most recent measurement from WMAP \cite{WMAP} has measured the basic parameters of cosmology to high precision. The density of dark energy is estimated to be $\Omega_{\Lambda 0} = 0.728^{+ 0.015}_{- 0.016}$. That is to say, about $72.8$ percent of the Universe consists of dark energy.

 We should remark that the measurements give the effective cosmological constant $\Lambda_{eff}$ which consists of two parts, the quantum zero point energy $\Lambda_{\mbox{dark energy}}$ and the geometric cosmological constant $\Lambda$, i.e.
$$\Lambda_{eff} = \Lambda_{\mbox{dark energy}} + \Lambda.$$

 It would be desirable to distinguish the contribution of $\Lambda$ from $\Lambda_{eff}$. We shall find that the OPERA experiment fits the need. The estimation of radius of the pseudo-sphere of de Sitter space gives an estimation of the geometric cosmological constant. Our estimation shows that the contribution of $\Lambda$ is about $10^{-4}$ compared with $\Lambda_{eff}$. So for the first time we are able to distinguish the contribution of the geometric cosmological constant and the quantum zero point energy. We shall explain the estimation in the following sections.

 In Section 2 we shall review Special Relativity with de Sitter spacetime symmetry. Such a Special Relativity was first realized by K. H. Look and his collaborators \cite{Lu74}. With its later developments \cite{Ours,Guo1,guo3,yan2}, e.g. it has a Lagrangian-Hamiltonian formalism \cite{Ours}, the theory can be adapted to our needs.

 In Section 3 we recall the explanation of OPERA experiment within Special Relativity with de Sitter spacetime symmetry. The experiment can be understood and as an outcome the radius of the pseudo-sphere can be estimated.

 In Section 4 we give an estimate of the geometric cosmological constant based on the estimation of radius.
 We can distinguish the contribution of the geometric cosmological constant and the quantum zero point energy
 from now.

\section{Review of Special Relativity with de Sitter spacetime symmetry (dS-SR)}

The Special Relativity is a theory on global spacetime geometry. The usual assumption of spacetime metric is the Minkowski metric $\eta_{\mu \nu} = diag\{1, -1, -1, -1\}$. Its characterizing property is that the general transformation to preserve the metric is the Poincare group (or the inhomogeneous Lorentz group $ISO(1, 3)$).
The Poincare group is the limit of the de Sitter group with radius $R \rightarrow \infty$. In the 1970's K. H. Look
and his collaborators Z. L. Zou and H. Y. Guo \cite{Lu74} proposed a Special Relativity with global de Sitter spacetime symmetry (dS-SR). Yan, Xiao, Huang and Li \cite{Ours} gave a Lagrangian-Hamiltonian formalism of dS-SR with two universal constants $c$ and $R$.

In Special Relativity with de Sitter spacetime symmetry we have the Beltrami metric $ds^{2} = \Sigma_{\mu \nu} B_{\mu \nu} dx^{\mu} \otimes dx^{\nu}$ of the global spacetime coordinates $x^{\mu}, \mu = 0, 1, 2,3$:

\begin{equation}\label{Beltrami}
B_{\mu \nu} = \frac{\eta_{\mu \nu}}{\sigma(x)} + \frac{\eta_{\mu \lambda} \eta_{\nu \rho} x^{\lambda} x^{\rho}}{R^{2} \sigma(x)^{2}}
\end{equation}

$$\sigma(x) := 1 - \frac{1}{R^{2}} \eta_{\mu \nu} x^{\mu} x^{\nu}$$

\noindent where the speed of light $c$ and the radius $R$ of the pseudo-sphere in the de Sitter space are universal constants.

The Lagrangian of a free particle in dS-SR reads

\begin{equation}\label{Lagrangian}
L_{dS} (t, x^{i}, \dot{x}^{i}) = - m_{0} c \sqrt{B_{\mu \nu}(x) \dot{x}^{\mu} \dot{x}^{\nu}}
\end{equation}

The equation of motion is

\begin{equation}\label{EOM}
v^{i} = \dot{x}^{i} = const
\end{equation}

Thus the coordinates for de Sitter spacetime can be used as an inertial frame metric.

In choosing spacetime coordinates we use the Big Bang (BB) as the natural origin. In the Earth
laboratory we are at the time $t_{0} = 13.7$Gy and $x_{0} = x(t_{0}) = 0$. The three dimensional
space is always isotropic and homogeneous by the Copernicus principle.

In the de Sitter spacetime we have ten parameters transformations preserving the Beltrami metric:

\begin{eqnarray}\label{transformation}
x^{\mu} \;-\hskip-0.10in\longrightarrow\hskip-0.3in^{dS} ~~~ \tilde{x}^{\mu}
&=& \pm \sigma(a)^{1/2} \sigma(a,x)^{-1} (x^{\nu}-a^{\nu})D_{\nu}^{\mu}, \\
\nonumber D_{\nu}^{\mu} &=& L_{\nu}^{\mu}+R^{-2} \eta_{\nu \rho}a^{\rho} a^{\lambda}
(\sigma(a) +\sigma^{1/2}(a))^{-1} L_{\lambda}^{\mu} , \\
\nonumber L : &=& (L_{\nu}^{\mu})\in SO(1,3), \\
\nonumber \sigma(x)&=& 1-\frac{1}{R^2}{\eta_{\mu \nu}x^{\mu} x^{\nu}}, \\
\nonumber \sigma(a,x)&=& 1-\frac{1}{R^2}{\eta_{\mu \nu}a^{\mu} x^{\nu}}.
\end{eqnarray}

It gives 10 conserved charges:

\begin{eqnarray}
\nonumber p_{dS}^i &=& m_0 \Gamma \dot{x}^i, \\
\nonumber E_{dS} &=&  m_0 c^2 \Gamma, \\
\label{physical momenta} K_{dS}^{i} & =& m_0 c \Gamma (x^i -t\dot{x}^i)=m_0c\Gamma x^i-tp_{dS}^i, \\
\nonumber L_{dS}^{i} & =& -m_0 \Gamma \epsilon^{i}_{\;jk} x^j \dot{x}^k=-\epsilon^{i}_{\;jk} x^jp^k_{dS}.
\end{eqnarray}

\noindent Here $E_{dS},{\mathbf p}_{dS},{\mathbf L}_{dS},{\mathbf K}_{dS}$ are
conserved physical energy, momentum, angular-momentum and boost
charges respectively, and $ \Gamma^{-1} $ is $\sigma(x) \frac{ds}{c dt}$.

It is straightforward to check the identity of $\sigma^2(x)B_{\mu\nu}(x)p_{dS}^\mu p_{dS}^\nu=m_0^2c^2$.
Then we have the dispersion relation for dS-SR \cite{Ours}
\begin{equation}\label{dp}
E_{dS}^2 =m_0^2 c^4+{\mathbf p}_{dS}^2 c^2 + \frac{c^2}{R^2}
({\mathbf L}_{dS}^2-{\mathbf K}_{dS}^2).
\end{equation}

When $R \rightarrow \infty$, above relation reduces to the usual dispersion relation of E-SR

$$E_{E}^2 =m_0^2 c^4+{\mathbf p}_{E}^2 c^2.$$

\section{The OPERA experiment of superluminal neutrinos from de Sitter Special Relativity}

The OPERA experiment can be understood in the framework of de Sitter Special Relativity.
We recall the derivation for the convenience of readers.

We consider neutrinos as free massive particles. Its speed can be derived through the dispersion relation.
Suppose the OPERA neutrinos moving trajectory is $\{x^{1} = x(t), x^{2} = 0, x^{3} = 0\}$, then
\begin{eqnarray}
\label{v def} v_{dS} &\equiv& \dot{x}(t)=\frac{c^2 p_{dS}}{E_{dS}}, \\
\label{energy} E_{dS} &=& \frac{m_0 c^2}{\sqrt{1-(\frac{v_{dS}}{c})^2+(\frac{x_0-v_{dS}t_0}{R})^2}},
\end{eqnarray}

\noindent where $t_0$ and $x_0$ are the OPERA neutrino moving's initial time and space location, i.e., $t_0\simeq 13.7Gy,\;\;x_0=x(t_0)\simeq 0$.

We have the formulae of neutrino velocity:

\begin{equation}\label{vdS}
v_{dS} = c \sqrt{\frac{1 - \frac{m_{0}^{2}c^{4}}{E^{2}}}{1 - \frac{c^{2}t_{0}^{2}}{R^{2}}}}
\end{equation}

We see when $E$ is large enough we could have $v_{dS} > c$.

In the OPERA experiment we have $E = 13.9 GeV$, $m_0 = 2 eV$ and

$$\delta v_{\nu} = \frac{v_{\nu} - c}{c} = (2.48 \pm 0.28_{stat} \pm 0.30_{sys}) \times 10^{-5}$$

We have:

$$\frac{v_{dS}}{c} = 1 + \delta v_{\nu} = 1 + \frac{1}{2} \frac{c^{2} t_{0}^{2}}{R^{2}} - \frac{1}{2} \frac{m_{0}^{2} c^{4}}{E^{2}} $$

By neglecting the higher order term we have:

$$\delta v_{dS} = \frac{1}{2} \frac{c^{2} t_{0}^{2}}{R^{2}}.$$

The radius $R$ of the pseudo-sphere can be estimated as

\begin{equation}\label{R}
R = \frac{c t_{0}}{\sqrt{2 \delta v_{dS}}} = (1.95 \pm 0.11 \pm 0.12) \times 10^{12} l.y.
\end{equation}

\section{Determination of the geometric cosmological constant from the OPERA superluminal experiment}

The estimation of $R$ has a deep consequence to cosmology. It gives an estimation of the geometric cosmological
constant $\Lambda$.

The Beltrami metric $B_{\mu \nu}$ satisfies Einstein's equation:

\begin{equation}\label{E eq}
{\cal R}_{\mu \nu} - \frac{1}{2} B_{\mu \nu} {\cal R} - \Lambda B_{\mu \nu} = 0.
\end{equation}

It is easy to see that for the Beltrami metric we have

$${\cal R}_{\mu \nu} = - \frac{3}{R^{2}} B_{\mu \nu}, {\cal R} = - \frac{12}{R^{2}}.$$

By substituting the above into Einstein's equation we have

\begin{equation}\label{Lambda}
\Lambda = \frac{3}{R^{2}}.
\end{equation}

Using $R = 1.95 \times 9.46 \times 10^{29} cm$ we get

\begin{equation}\label{Lambda num}
\Lambda \simeq 0.88 \times 10^{-60} cm^{-2}.
\end{equation}

We see that the observed cosmological constant $\Lambda_{eff}$ obey's the full Einstein's equation:

\begin{equation}\label{full E eq}
{\cal R}_{\mu \nu} - \frac{1}{2} g_{\mu \nu} {\cal R} - \Lambda g_{\mu \nu} = - \frac{8 \pi G}{c^{4}}(T_{\mu \nu} - \rho_{vac} c^2 g_{\mu \nu})
\end{equation}

\noindent where ${\cal R}$ is the scalar curvature, $G$ the Newton's gravitational constant, $\Lambda$ the geometric cosmological constant, $\rho_{vac}$ the matter's quantum zero point energy density( or $(T_{\mu \nu})_{vac} = - \rho_{vac} c^2 g_{\mu \nu})$ and $T_{\mu \nu}$ the energy-momentum tensor. And the effective cosmological constant $\Lambda_{eff}$ is:

\begin{equation}\label{Lambda eff}
\Lambda_{eff} = \frac{8 \pi G}{c^{2}} \rho_{vac} + \Lambda = \Lambda_{\mbox{dark energy}} + \Lambda.
\end{equation}

From measuring the effects of accelerate expansion of the Universe and the recent WMAP data \cite{WMAP} we have:

\begin{equation}\label{Lambda eff num}
\Lambda_{eff} = \frac{3 H_{0}^{2}}{c^{2}} \Omega_{\Lambda 0}
\simeq 1.26 \times 10^{-56} cm^{-2}.
\end{equation}

Finally we have:

\begin{equation}\label{Lambda dark}
\Lambda_{\mbox{dark energy}} = \Lambda_{eff} - \Lambda
= \Lambda_{eff} (1 - \frac{3}{R^2\Lambda_{eff}}) = \Lambda_{eff} (1 - O(10^{-4})),
\end{equation}
where Eqs.~(\ref{Lambda}) (\ref{Lambda num}) (\ref{Lambda eff num}) were used.
Therefore the geometric cosmological constant gives a small correction to the density of dark energy, which arises from the OPERA measurement.

\section{Summary and discussions}

The recent OPERA experiment of superluminal neutrinos is a breakthrough in physics. The experiment can be understood in the framework of de Sitter Special Relativity. And it has deep consequences to cosmology. The OPERA experiment can be viewed as an confirmation of the global spacetime geometry is de Sitter spacetime symmetry and the Poincare symmetry of Minkowski spacetime is an approximation. The radius of the pseudo-sphere of de Sitter spacetime can be estimated from the experiment.

In this paper we explore further consequences of the OPERA experiment to cosmology. The estimation of the radius can be used to determine the geometric cosmological constant $\Lambda$. From observations in cosmology one
can determine the effective cosmological constant $\Lambda_{eff}$ which is the sum of the quantum zero point energy
$\Lambda_{\mbox{dark energy}}$ and the geometric cosmological constant $\Lambda$. Thus it is the first time we have an estimation of the quantum zero point energy $\Lambda_{\mbox{dark energy}}$. We anticipate further estimations of $\Lambda_{\mbox{dark energy}}$ from $\Lambda_{eff}$ should adjust the contribution of the geometric cosmological constant.

\begin{center} {\bf ACKNOWLEDGMENTS}
\end{center}
  We would like to thank Professors Chen Bin, Kong Liang, Li Qin, Wang Shikun, Wu Ke and Yu Ming for helpful discussions. We would like to thank Professor Shing-Tung Yau for the invitation to the Second Tsinghua Sanya Mathematics Forum which is crucial to the work.
This work is partially supported by National Natural Science Foundation of China under Grant
No.~10975128 and No.~11031005 and by the Wu Wen-Tsun Key Laboratory of Mathematics at USTC of Chinese Academy of Sciences.

\end{document}